\documentclass[aps,prd,nofootinbib]{revtex4}
\usepackage[colorlinks=true, pdfstartview=FitV, linkcolor=blue, citecolor=red, urlcolor=magenta]{hyperref}
\usepackage{graphicx}
\usepackage{latexsym}
\usepackage{amsmath}
\usepackage{amsfonts}
\usepackage{amssymb}
\usepackage{verbatim}
\usepackage{float}
\begin{document}
\title{Dynamics for Spin-$1/2$ Particles in Einstein-Gauss-Bonnet Gravity}

\author{$^{1,2}$ E. Maciel}
\email{eugenio.maciel@df.ufcg.edu.br}

\affiliation{$^{1}$Unidade Acad\^emica de Engenharia de Produ\c{c}\~ ao, Universidade Federal de Campina Grande,\\
Caixa Postal 10071, 58540-000, Sum\'e, Para\'{\i}ba, Brazil.}

\affiliation{$^{2}$Unidade Acad\^emica de F\'{\i}sica, Universidade Federal de Campina Grande,\\
Caixa Postal 10071, 58429-900, Campina Grande, Para\'{\i}ba, Brazil.}

\begin{abstract}
In this work, I investigate the quantum dynamics of a spin-1/2 particle propagating in the spacetime of a static, spherically symmetric Einstein-Gauss-Bonnet (EGB) black hole within the framework of relativistic quantum mechanics. Starting from the Dirac equation in curved spacetime-formulated using the tetrad formalism and the associated spin connection we construct the corresponding Dirac Hamiltonian in EGB geometry. Next, I employ Heisenberg's equations of motion to derive explicit operator expressions for the particle's velocity and force, providing a fully quantum description of fermionic motion in a higher-curvature gravitational background. It is observed that the spacetime geometry modifies the Dirac dynamics through the EGB metric function, leading to corrections in the velocity and force operators that depend explicitly on the Gauss-Bonnet coupling parameter, $\xi$. In the weak-field limit, the expectation values of these operators satisfy Ehrenfest's theorem, demonstrating that the corresponding classical Einstein-Gauss-Bonnet dynamics emerges naturally as the semiclassical limit of the underlying quantum theory. In particular, the effective radial force includes higher-curvature contributions that become increasingly significant in the strong-gravity regime, while continuously reducing to the Schwarzschild result as the Gauss-Bonnet parameter approaches zero. These results establish a direct connection between relativistic quantum dynamics and modified gravity, providing an operator-based framework for investigating fermionic motion in Einstein-Gauss-Bonnet spacetimes.
\end{abstract}
 
\maketitle

\pretolerance10000
\section{Introduction}
\label{Intro}
In recent decades, the search for a unified framework capable of describing the four fundamental interactions of nature has remained one of the central goals of theoretical physics. In this context, gravity stands out as the most challenging interaction to incorporate consistently into a quantum description, due to its incompatibility with quantum field theory and the geometric nature of its most successful classical formulation, namely General Relativity. Einstein-Gauss-Bonnet (EGB) gravity constitutes one of the most natural higher-curvature extensions of general relativity. It arises as the second-order term in Lovelock theory, preserving second-order field equations while incorporating quadratic curvature corrections \cite{lovelock1971einstein}. In dimensions $D \geq 5$, the theory is well-defined and leads to nontrivial modifications of black hole solutions, such as the Boulware--Deser spacetime \cite{Boulware1985}. More recently, effective four-dimensional formulations of EGB gravity have been proposed and extensively discussed, although their consistency and physical interpretation require careful treatment \cite{glavan2020einstein,gurses2020there,Fernandes2022Review,Odintsov,Odintsov2,Odintsov3}. Within this effective framework, the Gauss--Bonnet coupling parameter $\xi$ encodes higher-curvature corrections, which become particularly relevant in regions of strong gravitational fields. Black hole solutions in Einstein-Gauss-Bonnet gravity have been widely investigated from the perspective of strong-field phenomenology. These studies include analyses of geodesic motion, innermost stable circular orbits (ISCO), photon spheres, quasi-normal modes, and accretion processes \cite{Konoplya2010,Doneva2018,Atamurotov2021ChargedSpinning,Konoplya:2020bxa,Konoplya:2020juj,Konoplya:2020ptx,Konoplya:2008ix}. A common feature of these analyses is that the Gauss--Bonnet parameter introduces corrections that become significant at short distances, providing a potential observational window for physics beyond Einstein gravity.

In parallel, fermionic fields propagating in EGB black-hole backgrounds have also been explored in the literature, primarily in the context of quasi-normal spectra and stability analyses \cite{Gonzalez2018FermionicGB,Churilova2021}. These works demonstrate that Dirac fields are sensitive to Gauss--Bonnet corrections, exhibiting both perturbative and non-perturbative responses. However, most of these studies rely on wave-equation approaches and focus on spectral properties, rather than on the underlying operator structure governing the quantum dynamics. On the other hand, the formulation of the Dirac equation in curved spacetime, based on the tetrad formalism and spin connection, provides a natural framework for describing fermionic dynamics in gravitational fields \cite{Brill1957,dewitt1966dynamical}. In this context, Hamiltonian formulations of the Dirac equation have been developed and extensively studied, particularly in static spacetimes such as Schwarzschild and Kerr geometries \cite{Parker1980,Obukhov2013,Silva2019}. These formulations establish a direct connection between relativistic quantum mechanics and operator dynamics. A particularly important feature of the Hamiltonian formulation is that it naturally provides the evolution of quantum observables through the Heisenberg equations of motion. In contrast to approaches based on wave equations or semiclassical trajectories, the Heisenberg picture assigns a well-defined operator to each observable, allowing quantities such as velocity, momentum and force to be derived directly from the Dirac Hamiltonian through commutator algebra. Consequently, the dynamics is formulated entirely at the operator level before any semiclassical approximation is introduced. Within this framework, physical observables such as position, momentum, and spin evolve according to their commutators with the Hamiltonian operator. This method has been successfully applied in various contexts, including relativistic quantum systems in external fields and curved spacetimes \cite{Obukhov2013,Hehl1990}. In gravitational backgrounds, the Heisenberg picture provides a direct route to defining velocity and force operators, thereby offering an operator-based interpretation of particle motion that goes beyond classical geodesic descriptions.

Despite these advances, the operator formulation of fermionic dynamics in Einstein-Gauss-Bonnet spacetimes remains largely unexplored. Existing studies have predominantly concentrated on wave propagation, quasi-normal modes, stability analyses, or classical geodesic motion, whereas the quantum dynamics of observables has received considerably less attention. In particular, to the best of our knowledge, an explicit derivation of the velocity and force operators from the Dirac Hamiltonian in an Einstein-Gauss-Bonnet background and the identification of how higher-curvature corrections modify these observables has not been presented so far. The primary objective of this work is to fill this gap. We construct the Dirac Hamiltonian for a spin-1/2 particle in a static, spherically symmetric Einstein-Gauss-Bonnet black hole spacetime using the tetrad formalism. From this Hamiltonian, we derive the Heisenberg equations of motion for the position and momentum operators, thereby obtaining explicit expressions for the velocity and force operators. This allows us to identify how the Gauss-Bonnet coupling modifies the effective fermionic dynamics at the operator level. This approach offers a perspective complementary to standard analyses based on geodesics and wave equations. 

Rather than focusing on classical trajectories or spectral properties, we obtain an operator-based description of motion that naturally incorporates both quantum and relativistic effects. In particular, we observe that the resulting force operator contains corrections proportional to the Gauss-Bonnet parameter, leading to a modified radial dependence that encodes higher-curvature effects in a manner directly comparable to classical forces. It is important to emphasize that the purpose of the present work is not simply to reproduce the weak-field Einstein-Gauss-Bonnet force. Instead, our starting point is the relativistic Dirac Hamiltonian, from which the velocity and force operators are obtained through the Heisenberg equations of motion. The classical dynamics emerges only after taking expectation values and invoking Ehrenfest’s theorem. Therefore, the weak-field force should be interpreted as the semiclassical limit of an underlying relativistic quantum dynamics rather than as the starting point of the analysis. Here, natural units ($\hbar=c=G=1$) will be used.

This paper is organized as follows. In Sec.~\ref{SC1}, we review the main aspects of Einstein--Gauss--Bonnet gravity and discuss the geometric structure of the corresponding spacetime. In Sec.~\ref{SC2}, we construct the Dirac Hamiltonian in this background and derive the associated velocity and force operators within the Heisenberg formalism. Finally, in Sec.~\ref{remarks}, we present our conclusions and discuss possible extensions of the present work.

\section{The Theory}
\label{SC1}
The starting point of our analysis is the Einstein-Gauss-Bonnet theory, which represents the second-order term in the Lovelock hierarchy \cite{lovelock1971einstein}. The corresponding action in $D$ dimensions space is given by
\begin{equation}
\label{TT1}
S_{\mathrm{EGB}}=\frac{1}{16 \pi G} \int d^D x \sqrt{-g}\left[R+\xi \mathcal{G}\right].
\end{equation}
where $R$ is the Ricci scalar, $\xi$ is the Gauss-Bonnet coupling parameter\footnote{In the literature, the standard notation for the Gauss-Bonnet parameter is $\alpha$. However, since we will be investigating this theory in the fermionic sector, we will explicitly consider the quantity $\alpha$ to be the Dirac matrix. Therefore, in this work, the Gauss-Bonnet parameter will be described by $\xi$.}. The quantity
\begin{equation}
\label{TT2}
\mathcal{G}=R_{\mu \nu \rho \sigma} R^{\mu \nu \rho \sigma}-4 R_{\mu \nu} R^{\mu \nu}+R^2.
\end{equation}
is the Gauss-Bonnet invariant defined in terms of the curvature tensor $R_{\mu\nu}$. It is well known that $\mathcal{G}$ is a topological term in four dimensions and does not contribute to the field equations when the action is treated in the standard way \cite{lovelock1971einstein}. Consequently, nontrivial dynamical effects from the Gauss–Bonnet term arise naturally only in dimensions $D>4$, as in the original formulation of Lovelock gravity. In ordinary four-dimensional spacetime, the quantity $\mathcal{G}$ is a topological invariant from the Gauss-Bonnet theorem and does not contribute to the classical equations of motion since its variation reduces to a boundary term \cite{lovelock1971einstein,zumino1986gravity}. 

In other words, the EGB theory, in its standard form, does not introduce new dynamical degrees of freedom in four dimensions. More recently, however, an effective four-dimensional formulation of EGB gravity has been proposed through a regularization procedure in which the coupling is rescaled as $\xi \to \xi/(D-4)$, followed by the limit $D \to 4$ taken at the level of the field equations \cite{glavan2020einstein}. Although this construction has triggered considerable debate regarding its fundamental consistency, several works have shown that the resulting theory can be consistently interpreted as an effective scalar–tensor model or as a well-defined limit of higher-dimensional Lovelock gravity see Refs. \cite{kumar2020gravitational,gross1987quartic,kanti1996dilatonic}. In the present work, we adopt this effective four-dimensional geometry as a phenomenological background spacetime. Our purpose is not to investigate the fundamental consistency of the four-dimensional EGB theory itself, but rather to study the dynamics of Dirac fermions propagating in the resulting effective geometry. Therefore, the metric is treated as a given classical background encoding higher-curvature corrections through the parameter $\xi$.

It is worth emphasizing that the effective four-dimensional metric contains no explicit dependence on the spacetime dimension. All higher-curvature information is encoded in the coupling parameter $\xi$, which parametrizes deviations from the Schwarzschild geometry. Consequently, observable quantities derived from this background depend only on the effective metric, while their connection with the underlying higher-dimensional theory remains implicit through the value of $\xi$. This feature makes the model particularly suitable for phenomenological investigations, since possible signatures of higher-curvature gravity may be explored entirely within a four-dimensional effective description.

\subsection{The Einstein-Gauss-Bonet Geometry}

Static and spherically symmetric black-hole solutions constitute the simplest and most extensively studied class of Einstein-Gauss-Bonnet geometries. \cite{Boulware1985}.  As emphasized in the previous section, although the Gauss–Bonnet invariant is topological in strictly four dimensions, effective formulations allow one to obtain nontrivial gravitational dynamics in $D=4$, encoding higher-curvature effects through the coupling parameter $\xi$. The metric takes the form
\begin{equation}
\label{TT3}
ds^2=-f(r)dt^2+\frac{dr^2}{f(r)}+r^2 d \Omega_{D-2}^2,
\end{equation}
where $d\Omega_{D-2}^2=\left(d\theta^{2}+\sin\theta d\phi^{2}\right)$ represents the metric of the $(D-2)$-dimensional sphere and the metric function $f(r)$ is the radial function given by
\begin{equation}
\label{TT4}
f(r)=1+\frac{r^2}{2 \xi}\left[1 \pm \sqrt{1+\frac{8 \xi M}{r^3}}\right].
\end{equation}
At small radial distances, the metric function $f(r)$ exhibits significant deviations from the Schwarzschild behavior due to the presence of higher-curvature contributions in Einstein-Gauss-Bonnet gravity. In particular, the square-root structure, introduces a nontrivial modification of the geometry as $r \to 0$. The argument of the square root remains positive for $\xi>0$, ensuring that the metric function is real-valued throughout the domain $r>0$. However, despite this formal regularity, curvature invariants such as the Kretschmann scalar still diverge at $r=0$, indicating that the central singularity is not removed within this effective four-dimensional framework. This behavior is consistent with previous analyses showing that the four-dimensional EGB solutions, although modifying the near-origin structure, do not generically resolve curvature singularities \cite{banerjee2021quark,jusufi2023charged}. 

Therefore, while the EGB corrections significantly alter the approach to the central region, they should be interpreted as modifying rather than eliminating the singular structure of the spacetime. The metric admits two distinct branches: The branch associated with the minus sign continuously reproduces the Schwarzschild solution in the limit $\xi\rightarrow0$, whereas the plus branch does not possess the general relativistic limit and is usually regarded as unstable. Here I am considering the first branch. Furthermore a fundamental requirement for any modified theory of gravity is the recovery of general relativity in the weak field regime. This is indeed satisfied in the present case. Expanding $f(r)$ to the first order in $\xi$ and taking $r$ large, one obtains
\begin{equation}
\label{TT5}
f(r) \simeq 1-\frac{2 M}{r}+\frac{4 \xi M^2}{r^4}.
\end{equation}
Unlike phenomenological modifications introduced directly at the level of the Hamiltonian, the corrections considered here originate entirely from the spacetime geometry. Consequently, every deviation from the Schwarzschild case possesses a clear geometric interpretation associated with higher-curvature effects. The first correction appears at order $1/r^4$, which is strongly suppressed at large distances. Therefore, the effects of the Gauss–Bonnet term become significant only in regions of high curvature, i.e., small $r$. Therefore, throughout this work we restrict ourselves to the first-order expansion in the Gauss–Bonnet parameter, which is sufficient to describe the leading higher-curvature corrections in the weak-field regime. The figure below clearly shows the effects of the EGB term on the usual behavior for general relativity.The figure clearly shows how the metric function $f(r)$ is modified by the inclusion of the Gauss–Bonnet term. 
\begin{figure}[H]
\centering
\includegraphics[scale=0.5]{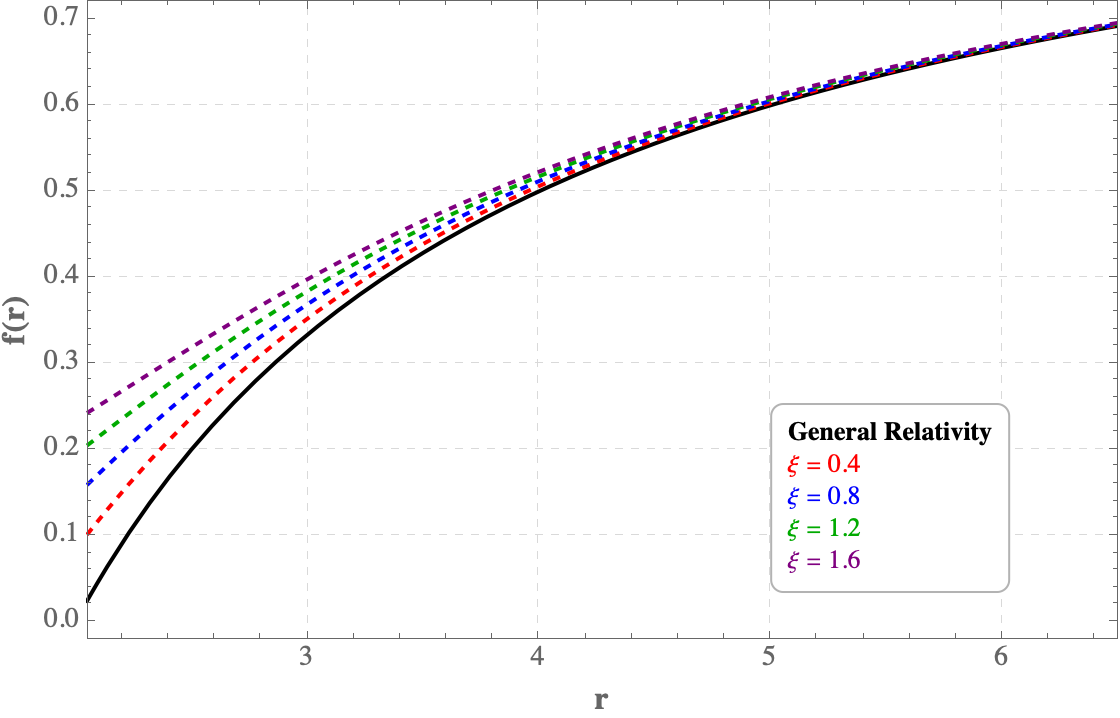}
\caption{
Metric function $f(r)$ as a function of the radial coordinate $r$, comparing the general relativity limit $\xi = 0$}
\end{figure}
The Figure 1 illustrates the influence of the Gauss-Bonnet parameter on the metric function. As expected, all curves converge to the Schwarzschild solution at large distances, confirming the recovery of General Relativity in the weak-field regime. On the other hand, noticeable deviations appear close to the event horizon, precisely where higher-curvature effects become relevant. Since the metric function enters directly into the Dirac Hamiltonian in curved spacetime, these geometric modifications propagate to the quantum dynamics investigated in the following sections. Furthermore, this effect can be analyzed in other aspects of black hole physics, given that the event horizon is defined by the condition $f(r_h)=0$. These corrections entail a shift in the horizon radius, altering fundamental properties such as Hawking temperature and entropy \cite{clunan2004gauss,banados1994black,estrada2023new,li2026entropic}. Another particularly relevant aspect is that, although explicit dependence on dimensionality does not appear in the final expression, the effects of extra dimensions remain encoded in the parameter $\xi$, such that the deviation between the curves can be interpreted as an indirect signature of high-energy physics.

\section{Dirac Fermions in Einstein-Gauss-Bonet Gravity}
\label{SC2}

The formulation of the Dirac equation in curved spacetime provides the fundamental framework for describing spin-$1/2$ particles interacting with gravitational fields \cite{sakurai1969advanced,bjorken1965relativistic,dirac1932relativistic}. In particular, considerable attention has been devoted to the study of fermionic fields in modified gravity backgrounds, including (EGB) gravity \cite{churilova2021quasinormal,zinhailo2019quasinormal}. We consider a static, spherically symmetric spacetime described by the line element (\ref{TT4}). To determine the Dirac equation in curved spacetime, we need the tatrada field $e^{a}_{\mu}$ since this quantity connects the coordinates of Minkowski spacetime with curved spacetime. The relation between both is described by the definition
\begin{equation}
\label{DF1}
g_{\mu\nu} = \eta_{ab}\, e^a_{\ \mu} e^b_{\ \nu},
\end{equation}
where $\eta_{ab}=\mathrm{diag}(-1,1,1,1)$ are the Minkowski metric. Here, the Latin indices describe the coordinates of flat space and the Greek indices those of curved space and the Dirac matrices are defined such that $\gamma^{\mu}=e^{\mu}_{a}\gamma^{a}$  satisfy the Clifford algebra $\{\gamma^\mu,\gamma^\nu\}=2g^{\mu\nu}$. Using the metric (\ref{TT4}) and the definition (\ref{DF1}) find
\begin{equation}
\label{DF2}
e_\mu^a=\left(\begin{array}{cccc}
\sqrt{f} & 0 & 0 & 0 \\
0 & \frac{1}{\sqrt{f}} & 0 & 0 \\
0 & 0 & r & 0 \\
0 & 0 & 0 & r \sin \theta
\end{array}\right)\quad\mbox{,}\quad
e_a^\mu=\left(\begin{array}{cccc}
\frac{1}{\sqrt{f}} & 0 & 0 & 0 \\
0 & \sqrt{f} & 0 & 0 \\
0 & 0 & \frac{1}{r} & 0 \\
0 & 0 & 0 & \frac{1}{r \sin \theta}
\end{array}\right),
\end{equation}
and Dirac matrices are explicitly written as
\begin{equation}
\label{DF3}
\gamma^t=\frac{1}{\sqrt{f}} \gamma^0\quad\mbox{,}\quad
\gamma^r=\sqrt{f} \gamma^1\quad\mbox{,}\quad
\gamma^\theta=\frac{1}{r} \gamma^2\quad\mbox{,}\quad
\gamma^\phi=\frac{1}{r \sin \theta} \gamma^3.
\end{equation}

To determine the Dirac equation in curved space, it must be taken into account that the matrices of the transformation group are functions of the point; in this sense, the spinor derivative does not transform as a spinor \cite{collas2019dirac}. Thus, it is convenient to replace the ordinary derivative with the covariant derivative $\partial_{\mu}\rightarrow\nabla_{\mu}=\partial_\mu+\Gamma_{\mu}$where the quantity $\Gamma_{\mu}$ is the spinorial connection
\begin{equation}
\label{DF4}
\Gamma_\mu=\frac{1}{8}\,\omega_\mu^{ab}\left[\gamma_a,\gamma_b\right],
\end{equation}
and $\omega_\mu^{a b}=e^{a \nu}\left(\partial_\mu e_\nu^b-\Gamma_{\mu \nu}^\lambda e_\lambda^b\right)$ are the spin connection coefficients and $\Gamma_{\mu \nu}^\lambda$ are the Christoffel symbols. Thus, the Dirac equation for curved spacetime is
\begin{equation}
\label{DF5}
\Big[i\gamma^\mu\left(\partial_\mu + \Gamma_\mu\right)-m\Big]\psi=0.
\end{equation}
For our system described by the metric (\ref{TT3}) and taking into account the quantities from (\ref{DF1}) to (\ref{DF5}) we explicitly have the Dirac equation as being
\begin{align}
\label{DF6}
\Bigg[
\frac{i}{\sqrt{f}}\gamma^0 \partial_t
+ i\sqrt{f}\gamma^1\left(\partial_r+\frac{1}{r}+\frac{f'}{4f}\right)
+ \frac{i}{r}\gamma^2\left(\partial_\theta+\frac{\cot\theta}{2}\right)
+ \frac{i}{r\sin\theta}\gamma^3 \partial_\phi
- m
\Bigg]\psi = 0.
\end{align}
Since the metric function completely determines the tetrad fields and the associated spin connection, every higher-curvature correction encoded in $f(r)$ propagates consistently into the covariant Dirac equation. Consequently, the Einstein-Gauss-Bonnet parameter modifies the fermionic dynamics not through an additional interaction term, but through the geometry itself, affecting the spin connection and, ultimately, the relativistic Hamiltonian. From the equation above, it is possible to determine the Hamiltonian equation for a Dirac particle in EGB spacetime. For this, we need the standard structure $i\partial_{t}\psi=\mathcal{H}\psi$. Thus, multiplying the entire equation (\ref{DF6}) by $\gamma^{0}\sqrt{f}$ and isolating the time part, we find
\begin{align}
\label{DF7}
\mathcal{H} = 
\beta m \sqrt{f}
- i\alpha^1 f\left(\partial_r + \frac{1}{r} + \frac{f'}{4f}\right)
- i\sqrt{f}\alpha^2 \frac{1}{r}\left(\partial_\theta + \frac{\cot\theta}{2}\right)
- i\sqrt{f}\alpha^3 \frac{1}{r\sin\theta}\partial_\phi,
\end{align}
with $f^{\prime}=\partial_{r}f$. Here we introduce Dirac matrices in standard representation $\gamma^{0}=\beta$ and $\alpha^{i}=\gamma^{0}\gamma^{i}$. This Hamiltonian encodes the interaction between the fermionic degrees of freedom and the gravitational background, with the Gauss-Bonnet corrections entering explicitly through the metric function $f(r)$ and its radial derivative $f'(r)$.

\subsection{Heisenberg Dynamics: Velocity and Force Operators}

In quantum mechanics, the concept of force does not exist, since in this theory a particle does not follow a trajectory. Thus, when we say we are describing the dynamics of a particle from the point of view of relativistic quantum mechanics, we are referring to the determination of the equations for the velocity and force operators from Heisenberg's equations \cite{sakurai1969advanced,bjorken1965relativistic,dirac1932relativistic}. In fact, the correspondence principle, together with Ehrenfest's theme, guarantees an analogy between the results of quantum mechanics and those of classical mechanics, thus allowing us to describe its dynamics.\footnote{It will be quite common throughout the text to treat quantities only as velocity and force, for example, but it is important to highlight that we are actually dealing with operators. For this reason, in order not to overload the notation, we are disregarding in our work the notation ``$(\ \hat{}\ )$" on quantities, quite common to describe operators.} The Hamiltonian formulation derived in the previous section provides a natural framework for analyzing quantum dynamics in the Heisenberg picture. In this approach, the time evolution of any operator $\mathcal{A}$ is governed by
\begin{equation}
\label{DF8}
\frac{d\mathcal{A}}{dt} = i[H,\mathcal{A}],
\end{equation}
for time-independent operators. In particular, this allows us to define the velocity and force operators directly from the Dirac Hamiltonian, providing an operator-based description of particle motion in (EGB) spacetime. In the present work, I restrict the analysis of the velocity and force operators to their radial components. This choice is not merely a technical simplification, but rather a physically well-motivated and symmetry-driven reduction that captures the essential dynamical content of the problem.

First, it is important to emphasize that the background spacetime under consideration is static and spherically symmetric (\ref{TT3}). As a consequence of this symmetry, all nontrivial gravitational information is encoded in the radial dependence of the metric function $f(r)$. In particular, there are no intrinsic angular gradients in the geometry, so that the effective gravitational interaction is governed entirely by radial derivatives such as $\partial_r f(r)$. This implies that the force induced by the spacetime curvature is naturally central, pointing along the radial direction, in close analogy with Newtonian gravity and its relativistic generalization. Therefore, the radial component of the force operator directly captures the physically relevant gravitational interaction \cite{obukhov2001spin,ahluwalia1996gravitationally}. A second, independent justification arises from the structure of the Dirac equation in curved spacetime. In spherically symmetric backgrounds, the Dirac equation admits a separation of variables into radial and angular sectors, with the angular dependence described by spinor spherical harmonics, while the nontrivial dynamics is governed by radial differential equations \cite{kanti2004black}. As a result, the angular components of observables such as the velocity operator are largely determined by conserved quantities associated with rotational symmetry, whereas the radial component encodes the response of the fermionic field to the gravitational background. Consequently, focusing on the radial sector allows one to isolate the genuine dynamical effects induced by the geometry, without contamination from purely kinematical angular contributions.

We begin with the position operator $\mathbf{r}$ such that $r=\sqrt{x^{2}+y^{2}+z^{2}}$. In this sense, the velocity operator is by definition
\begin{equation}
\label{DF9}
 \frac{d {\bf{r}}}{dt} = i[\mathcal{H},{\bf{r}}]={\bf{v}}.
\end{equation}
Using the Hamiltonian (\ref{DF7}), the only non-vanishing contribution to the commutator arises from the momentum-dependent terms. After straightforward computation, we obtain
\begin{equation}
\label{DF10}
v(r) = f(r)\,\alpha,
\end{equation}
This expressions generalize the flat-spacetime result ${\bf v}= \alpha$ by incorporating the effects of spacetime curvature. In particular, the radial velocity operator acquires a multiplicative factor $f(r)$, reflecting the gravitational redshift and the influence of the Gauss-Bonnet parameter through the metric function. Expanding in the weak-field regime  (\ref{TT5}),we obtain
\begin{equation}
\label{DF11}
v(r) \simeq \alpha\left(1 - \frac{2M}{r} + \frac{4\xi M^2}{r^4}\right).
\end{equation}
From a physical standpoint, this result suggests that fermionic dynamics in an EGB background can be interpreted as those of a relativistic particle whose effective velocity is modulated by a scalar function that incorporates both the usual gravitational potential and high curvature corrections. Specifically, the presence of the term $\sim \xi M^2/r^4$ indicates that modified gravity effects manifest directly in the particle's local kinematics, and not just in derived quantities like the force. 

Another relevant aspect is that the operator's structure remains diagonal in the Dirac basis (matrix $\alpha$), which implies that the corrections do not introduce new spin couplings at this level, but rather deform the ``geometric environment'' in which the fermion propagates. This reinforces the interpretation that the Gauss–Bonnet term acts as a modification of the effective spacetime, and not as a new fundamental interaction in the fermionic sector. Finally, from a phenomenological point of view, this type of correction opens the possibility of accessing modified gravity effects through kinematic observables. In particular, explicit radial dependence can, in principle, lead to small changes in trajectories, characteristic frequencies, or even energy levels in bound systems, suggesting a possible path for the experimental investigation of the parameter $\xi$ in high-precision regimes \cite{Safronova2018,ALPHA2017_1S2S,ALPHA2017_spectroscopy,Gabrielse2006,KosteleckyRussell2011,maciel2025gravitational}.

Now, let's look at the force operator. Once again, we start from the Hamiltonian (\ref{DF7}) using Heisenberg's equation the force operator is defined through the time evolution of the momentum operator ${\bf p}=m{\bf v}$
\begin{equation}
\label{DF12}
\frac{d{\bf p}}{dt}=i[\mathcal{H},{\bf p}]={\bf F}
\end{equation}
In contrast to the velocity operator, the force operator exhibits a more direct sensitivity to the geometry, as it depends explicitly on derivatives of the metric function. In particular, one finds that $F_r \propto \partial_r f(r)+\operatorname{spin}-\text { dependent terms. }$Thus, the radial component of the force is
\begin{equation}
\label{DF13}
F(r) =
- \frac{mM}{r^2}
+ \frac{8m\xi M^2}{r^5}.
\end{equation}
This expression provides a particularly revealing framework, as it allows for the explicit separation of the usual gravitational contribution from the correction caused by Einstein-Gauss-Bonnet gravity. The first term in (\ref{DF13}) corresponds exactly to the effective Newtonian gravitational force, consistent with the expected behavior of General Relativity in the weak-field regime. This term describes a central attractive interaction, dominant at large distances. The dependence on $\sim 1/{r^5}$ for the EGB correction indicates that this contribution is strongly suppressed for $r \gg (M\xi)^{1/3}$, becoming relevant only in regions close to the gravitational source. 

This is consistent with the fact that the Gauss-Bonnet term involves quadratic curvature invariants, whose effects intensify in strong-field regimes. Moreover, the structure of the force operator suggests a close analogy with effective forces in external fields, with the metric function playing the role of a gravitational potential. This analogy provides a useful bridge between classical intuition and quantum dynamics. This point will be further elaborated in the section below.
\subsection{Classical Limit and Ehrenfest’s Theorem}

The operator equations obtained in the previous section describe the exact quantum dynamics of a Dirac particle propagating in the Einstein-Gauss-Bonnet background. In order to establish the connection with the corresponding classical motion, it is instructive to analyze the expectation values of these operators through Ehrenfest’s theorem \cite{sakurai1969advanced,bjorken1965relativistic, dirac1932relativistic}. For an arbitrary observable $\mathcal{A}$, the Heisenberg equation implies
\begin{equation}
\frac{d}{d t}\langle\mathcal{A}\rangle=\frac{i}{\hbar}\langle[H, \mathcal{A}]\rangle+\left\langle\frac{\partial \mathcal{A}}{\partial t}\right\rangle.
\end{equation}
Since both the position and momentum operators possess no explicit time dependence, their expectation values satisfy
\begin{equation}
\frac{d}{d t}\langle\mathbf{r}\rangle=\langle\mathbf{v}\rangle\quad\mbox{,}\quad
\frac{d}{d t}\langle\mathbf{p}\rangle=\langle\mathbf{F}\rangle.
\end{equation}
Here, the velocity and force operators derived from the Dirac Hamiltonian determine the evolution of the average position and momentum of the fermionic wave packet. In the weak field regime, where the wave packet remains sufficiently localized compared to the characteristic curvature scale of spacetime, the expected values of these quantities are approximated by
\begin{equation}
\left\langle\widehat{v}_r\right\rangle \simeq\left\langle\alpha^1\left(1-\frac{2 M}{r}+\frac{4 \xi M^2}{r^4}\right)\right\rangle\quad\mbox{,}\quad
\left\langle\hat{F}_r\right\rangle \simeq-\frac{2 M m}{r^2}+\frac{16 \xi M^2 m}{r^5}.
\end{equation}

It is worth emphasizing that the operator equations constitute the primary result of the present analysis. The classical force is recovered only after taking expectation values and invoking Ehrenfest’s theorem. Thus, although the final weak-field expression resembles the Newtonian limit corrected by Einstein-Gauss-Bonnet gravity, its derivation is fundamentally different: the dynamics follows directly from the quantum commutator algebra of the Dirac Hamiltonian rather than from the classical geodesic equation or from an effective gravitational potential. This result provides an important consistency check of the formalism, since it demonstrates that the Heisenberg equations reproduce the expected classical dynamics while preserving the complete operator structure characteristic of relativistic quantum mechanics. Although the expectation values reproduce the classical Einstein-Gauss-Bonnet dynamics in the weak-field limit, the underlying operator equations retain their fully relativistic and quantum character. Therefore, the agreement with the classical force should be interpreted as the correct semiclassical limit of the theory rather than as evidence that the operator derivation merely reproduces the classical potential.

\subsection{Discussions}
At first sight, the appearance of a Newtonian-like term in the force expression may seem surprising, since the starting point of the analysis is a fully relativistic Dirac Hamiltonian. However, this result is in fact expected and required for consistency. In the weak-field regime, the metric component $g_{tt}$ can be written as $g_{tt} \simeq -(1 + 2\Phi)$, where $\Phi$ is the effective gravitational potential. Using the expansion of the metric function, one recovers $\Phi(r) = -M/r + 2\xi M^2/r^4$, which immediately leads to a force of the form $F = -m \nabla \Phi$. Therefore, the Newtonian term emerges naturally as the leading contribution of the relativistic theory in the appropriate limit, while the Gauss–Bonnet correction appears as a higher-order modification. This demonstrates that the Dirac-based approach consistently reproduces the expected classical behavior while incorporating new physics associated with higher-curvature terms \cite{Parker1980,obukhov2001spin,JentschuraNoble2013}. This provides a new perspective on how quantum particles probe deviations from general relativity at the operator level. The figure below shows the behavior of the radial force (\ref{DF13}) for different values of $\xi$.
\begin{figure}[H]
\centering
\includegraphics[scale=0.5]{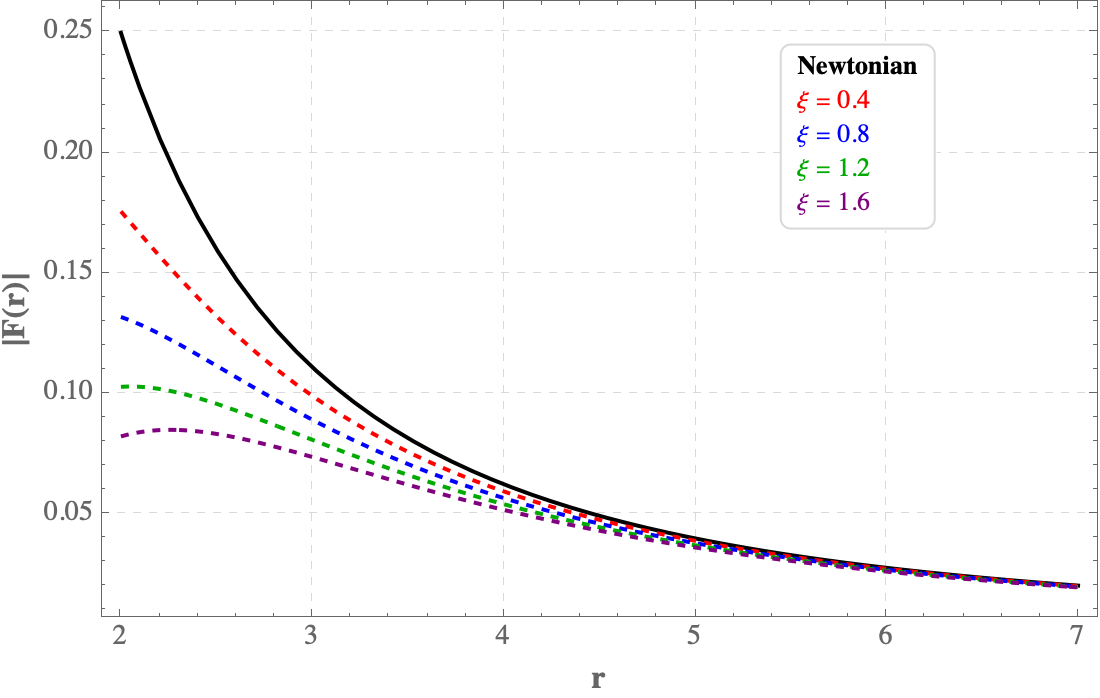}
\caption{Radial gravitational force as a function of $r$, comparing the newtonian limit with the EGB corrected profiles for different values of $\xi$}
\end{figure}

The figure quantitatively demonstrates the impact of corrections associated with Einstein–Gauss–Bonnet gravity on the radial gravitational force. While the solid curve corresponding to the General Relativity limit $\xi=0$ reproduces the standard behavior $F(r)\propto 1/r^2$, the dashed curves show that the introduction of a parameter $\xi \neq 0$ leads to a systematic modification of the intensity of the gravitational interaction, particularly pronounced in the region of small distances. It is observed that, as $\xi$ increases, the effective force undergoes a relative reduction compared to the usual case, indicating that the quadratic curvature terms act to ``\textit{smooth}'' the gravitational field in high curvature regimes. This behavior is consistent with the fact that Gauss–Bonnet theory introduces relevant corrections in the strong field regime, while preserving the asymptotic limit of General Relativity for large $r$, as demonstrated \cite{zinhailo2019quasinormal,churilova2021quasinormal}. From a phenomenological point of view, the separation between the curves provides a direct measure of the sensitivity of the gravitational force to the parameter $\xi$, suggesting that observables associated with strong field regimes may impose non-trivial constraints on the magnitude of these corrections. 

Furthermore, this result acquires an additional degree of relevance when considering that it emerges in the context of a fermionic theory, in which the dynamics of spin-1/2 particles is described from an effective Hamiltonian derived from the Dirac equation in an Einstein–Gauss–Bonnet gravity background. Unlike most studies in the literature, which focus on purely geometric solutions or classical aspects of the theory, the present analysis explicitly incorporates the effects of Gauss–Bonnet gravity on the quantum dynamics of fermions, allowing direct access to physical observables such as the force operator in the Heisenberg frame and thus comparing them with classical results through the correspondence principle. The central novelty, therefore, lies in establishing an explicit connection between higher-order curvature terms and the dynamical response of quantum particles, something that goes beyond the traditional description based on geodesics or classical effective potentials. In this sense, the systematic deviation observed in the curves from the limit of General Relativity is not merely a geometric consequence, but reflects modifications in the very structure of the fermionic Hamiltonian, with possible implications in systems where quantum and gravitational effects coexist. Therefore, the present formalism demonstrates that higher-curvature corrections are encoded directly in the operator dynamics generated by the Dirac Hamiltonian. The agreement with the classical Einstein–Gauss–Bonnet force is recovered only after taking expectation values and invoking Ehrenfest’s theorem, showing that the classical dynamics emerges naturally as the semiclassical limit of the relativistic quantum theory rather than being assumed from the outset. Furthermore, the fact that these corrections are parameterized by $\xi$ and manifest themselves measurably in the effective force suggests a promising path for the phenomenological investigation of the theory, especially in scenarios where the dynamics of Dirac particles in intense gravitational fields can be explored. Thus, the result not only broadens the scope of applications of Gauss–Bonnet gravity, but also introduces a new perspective by consistently integrating it into the domain of fermionic particle physics.

\section{Final Remarks}
\label{remarks}
In this work, we have investigated the dynamics of spin-$1/2$ particles in the background of a four-dimensional Einstein-Gauss-Bonnet (EGB) spacetime, with particular emphasis on the construction and interpretation of the velocity and force operators in the Heisenberg picture. Starting from the Dirac Hamiltonian in curved spacetime, we derived explicit expressions for these operators and analyzed how higher-curvature corrections modify fermionic motion. A central aspect of our approach is the adoption of an effective four-dimensional formulation of Einstein--Gauss--Bonnet gravity. Although the Gauss--Bonnet term is topological in strictly four dimensions \cite{lovelock1971einstein}, effective constructions allow one to encode higher-curvature effects through the coupling parameter $\xi$ \cite{glavan2020einstein,Casadio2020}. Within this framework, we have shown that the spacetime geometry is fully characterized by a modified metric function $f(r)$
provides a clear separation between the standard Schwarzschild contribution and the leading Gauss-Bonnet correction \cite{Boulware1985}. A key result of this work is that the influence of the Gauss--Bonnet term on fermionic dynamics is primarily encoded in the radial structure of the geometry. This justifies the restriction of our analysis to the radial components of the velocity and force operators, which capture the genuine interaction between the fermionic field and the gravitational background. Within this sector, we demonstrated that the velocity operator is modified through a deformation of the effective coupling to the metric, while the force operator acquires a distinct correction scaling as $1/r^5$, directly associated with the higher-curvature term. From a physical perspective, these results show that Einstein--Gauss-Bonnet gravity does not introduce new anisotropic interactions at the level of fermionic dynamics, but rather modifies the strength and radial profile of the central gravitational field. The correction governed by $\xi$ is subleading at large distances, ensuring consistency with general relativity, but becomes increasingly relevant in regimes of high curvature.

An important conceptual contribution of this work lies in establishing a direct link between modified gravity and fermionic observables. By working at the level of the Dirac Hamiltonian and the associated operator dynamics, we have shown that geometric corrections can be translated into measurable effects on quantities such as velocity and force. This provides a complementary perspective to traditional analyses based on geodesics or classical potentials \cite{Parker1980,obukhov2001spin}.
From a phenomenological standpoint, the presence of well-defined corrections suggests that constraints on the Gauss--Bonnet parameter $\xi$ could, in principle, be obtained from high-precision experiments. In particular, atomic spectroscopy and precision measurements in quantum systems have been widely recognized as powerful probes of new physics \cite{Safronova2018}, while experiments with antihydrogen \cite{ALPHA2017_spectroscopy} and Penning traps \cite{Gabrielse2006} provide complementary platforms sensitive to small deviations in dynamical observables. Finally, the framework developed here opens several directions for future investigation. These include the extension of the analysis to non-static or non-spherically symmetric backgrounds, the study of bound states and energy spectra in the presence of Gauss--Bonnet corrections, and the exploration of possible experimental signatures in high-precision quantum systems. In this sense, the present work provides a consistent and physically transparent foundation for further studies at the interface between modified gravity and quantum dynamics.

{\acknowledgments} EM thanks to the Graduate Program in Physics at the Federal University of Campina Grande PPGF.

\end{document}